# Reversible, Opto-Mechanically Induced Spin-Switching in a Nanoribbon-Spiropyran Hybrid Material

Bryan M. Wong,[1,]* Simon H. Ye,[2] and Greg O'Bryan[1]

[1]*Materials Chemistry Department, Sandia National Laboratories, Livermore, California 94551, USA*

[2]*Department of Chemistry, Stanford University, Stanford, California 94309, USA*

*Corresponding author: bmwong@sandia.gov

Web: http://alum.mit.edu/www/usagi

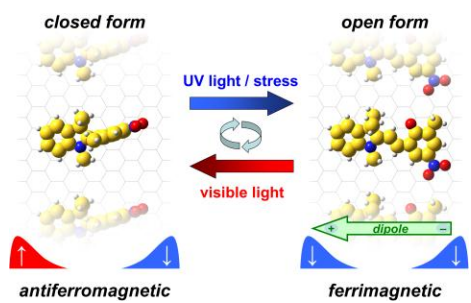

**Caption:** First-principles calculations show that the spin distribution in a nanoribbon-spiropyran hybrid material can be *reversibly* modulated via optical and mechanical stimuli without the need for large external electric fields.

**Abstract**

It has recently been shown that electronic transport in zigzag graphene nanoribbons becomes spin-polarized upon application of an electric field across the nanoribbon width. However, the electric fields required to experimentally induce this magnetic state are typically large and difficult to apply in practice. Here, using both first-principles density functional theory (DFT) and time-dependent DFT, we show that a new spiropyran-based, mechanochromic polymer *noncovalently* deposited on a nanoribbon can collectively function as a dual opto-mechanical switch for modulating its own spin-polarization. These calculations demonstrate that upon mechanical stress or photoabsorption, the spiropyran chromophore isomerizes from a closed-configuration ground-state to a zwitterionic excited-state, resulting in a large change in dipole moment that alters the electrostatic environment of the nanoribbon. We show that the electronic spin-distribution in the nanoribbon-spiropyran hybrid material can be *reversibly* modulated via noninvasive optical and mechanical stimuli without the need for large external electric fields. Our results suggest that the reversible spintronic properties inherent to the nanoribbon-spiropyran material allow the possibility of using this hybrid structure as a resettable, molecular-logic quantum sensor where opto-mechanical stimuli are used as inputs and the spin-polarized current induced in the nanoribbon substrate is the measured output.

**Introduction**

The quantum control of spin-polarization effects in nanomaterials via external stimuli is a necessary ingredient towards next-generation spintronic devices for information storage,[1-3] optoelectronic sensors,[4-6] and efficient energy conversion.[7-9] Organic nanomaterials are appealing as promising building blocks in this arena due to intrinsic weak spin-orbit and hyperfine interactions that allow electronic spin polarization to persist over larger distances compared to conventional semiconductors.[10,11] As a result, much work in this area has focused on zigzag graphene nanoribbons (ZGNRs) that intrinsically possess ferromagnetically-ordered electronic states along each ribbon edge.[12-14] In order to tune the spintronic properties of these nanomaterials, Son, Cohen, and Louie have demonstrated that ZGNRs can become half-metallic when an external electric field is applied across the nanoribbon width.[13] Achieving this half-metallicity is of great interest in spintronic applications since these materials have a large band gap for one spin type, while the energy bands for the opposite spin show metallic character (i.e., a zero band gap), resulting in a higher conductance of only one spin type through the material.[10,15]

Although the use of external electric fields offers a user-tunable procedure for obtaining spin-polarized transport, the field strength required to induce these effects experimentally is quite large and difficult to apply in practice. In principle, it is more realistic to design composite ZGNR-hybrid materials that already demonstrate inherent half-metallic behavior. Motivated by this possibility, several other researchers have proposed various different methods for modulating spin-polarization in ZGNRs, such as electronic doping with nitrogen or boron atoms,[16-18] direct functionalization by forming chemical bonds with the ZGNR sheet/edge,[19,20] or noncovalent physisorption on the ZGNR surface.[21-23] Among these three different methods, the first two involve the direct introduction of charge carriers in the material that ultimately destroy the electronic structure of the original ZGNR. In contrast, noncovalent physisorption (which is inherently driven by favorable molecular self-assembly) involves dispersive interactions that do not significantly modify band structure properties, leaving the unique electronic structure of the parent ZGNR intact.

Herein, we demonstrate that a spiropyran-embedded polymer noncovalently deposited on a ZGNR can intrinsically function as a dual opto-mechanical switch for modulating its own spin-selectivity. In previous experimental and theoretical work, we have shown that a di-functional spiropyran can be incorporated within a polymer backbone, lending itself towards both photochromic and mechanochromic activation.[24] Upon UV photoexcitation or application of mechanical stress, the closed spiropyran unit readily isomerizes to an open, charge-separated merocyanine state, as shown in Figure 1. This isomerization process is reversible upon irradiation with visible light, which restores the original closed spiropyran state. Using both first-principles density functional theory (DFT) and time-dependent DFT, we demonstrate that the open merocyanine form has a sizeable dipole moment that can be harnessed to generate a potential drop across the ZGNR substrate, effectively breaking its spin symmetry. Furthermore, since this isomerization process is reversible, the nanoribbon-spiropyran hybrid system can effectively function as a resettable molecular-logic system by using opto-mechanical stimuli as inputs and the spin-polarized current in the hybrid structure as the output. Our first-principles calculations give detailed insight into these electronic mechanisms and predict the specific opto-mechanical conditions that induce this spin-symmetry breaking in our nanoribbon-spiropyran hybrid system. Finally, we conclude our investigation by describing a number of realistic applications for further experiments, and we discuss possible modifications of the nanoribbon-spiropyran structure for inducing larger spin-polarization effects in this hybrid system.

**Results and Discussion**

We commence our investigation by calculating equilibrium geometries and binding energies of the various nanoribbon-spiropyran hybrid structures. Since the spiropyran chromophore reversibly isomerizes between two stable forms in an embedded polymer, we carried out geometry optimizations for both of these configurations noncovalently deposited on graphene nanoribbons. In order to quantitatively predict the π-π stacking interactions between these different materials, we carried out spin-polarized calculations with a dispersion-corrected Perdew-Burke-Ernzerhof (PBE-D) functional[25,26] using ultrasoft pseudopotentials and one-dimensional periodic boundary conditions along the axis of the nanoribbon (further details are given in the Theoretical Methods section). Since the size (as measured along the *x*-axis in Figure 2) of a spiropyran molecule is over 3 times longer than a ZGNR unit cell, a large supercell of 9.8 Å along the periodic *x*-axis of the nanoribbon was chosen (see Figure 3), which allows a separation distance of ~ 5 Å between adjacent spiropyrans. This separation distance is a reasonable choice since the individual spiropyran units lie in close proximity to each other within the experimental, spiropyran-embedded polymer.[24] In all of our calculations, we used a free-standing ZGNR containing 8 units of the C=C pair along the *y*-direction per unit cell. An extremely large vacuum space of 100 Bohrs along the *y*- and *z*-axes was imposed on our fully-periodic calculations in order to avoid spurious dipole interactions between adjacent spiropyrans in repeated supercells (using this large vacuum space, we also found dipole corrections to be negligible).

Figures 2 and 3 depict our converged equilibrium geometries for the closed and open spiropyran forms deposited on ZGNR. After several unconstrained combinatorial geometry optimizations of the ions and supercell, we found that both the closed and open spiropyran forms lie relatively flat across the ZGNR plane with a non-bonded separation distance of 2.4 Å. These distances and orientations are in accord with a recent theoretical study where a stacking distance of 2.5 Å was obtained for polar PVDF polymers on ZGNR.[23] Similar to other studies on dipolar interactions,[27,28] we also found that our composite structures were stabilized when the dipole alignment of the polymer was oriented along the width of the nanoribbon. As discussed in Ref.

23, this particular dipole orientation is favored due to a weakly-induced image dipole in the low-screening semi-conducting ZGNR. It is also important to point out that the orientations of both the closed and open forms as shown in Figures 2 and 3 correspond to an experimental setup where the spiropyran-embedded polymer chains are oriented across the nanoribbon width (i.e. the polymer is stretched along the $y$-axis; cf. Figure 2 in Ref. 24), in contrast to previous work on PVDF chains, which lie parallel to the ZGNR axis.[23]

Based on our equilibrium geometries, we calculated favorable binding energies ($E_b$ > 0.5 eV per spiropyran molecule) for both the closed and open nanoribbon-spiropyran structures, indicating the noncovalent interactions that stabilize these hybrid structures are energetically favorable. To provide further insight into their electronic properties, we calculated spin-polarized band structures for both an isolated ZGNR and for the optimized closed and open nanoribbon-spiropyran hybrid structures. In our calculations for the unmodified ZGNR, our PBE-D calculations confirm that the antiferromagnetic spin configuration (with up- and down-spin states localized on opposite zigzag edges – see Figure 4) is the ground state and is significantly more stable than the ferromagnetic state. We also note that a plot of the spin density of the ZGNR looks similar to Figure 4, with spins on each edge being ferromagnetically coupled (see, for example, Fig. 1(a) in Ref. 29). Since the edge configurations in a pristine ZGNR are identical, the up- and down-spin states are energetically degenerate, and as shown in Figure 5, the unmodified ZGNR has identical band structures for both spins with a band gap of 0.45 eV. When a closed spiroypyran is noncovalently deposited on ZGNR, the up- and down-spin states are poised for symmetry breaking, and a very slight band re-ordering can be seen in the overall band structure. However, upon further mechanical stress or UV photoexcitation, the closed spiropyran isomerizes to a charge-separated zwitterionic state that has sizeable dipole moment of 11.5 Debye (discussed further in the following paragraphs). The resulting dipole creates a substantial electric potential energy difference that is raised near the negatively-charged C=O group and lowered at the positively-charged nitrogen atom (cf. Figure 1). Since the open-spiropyran configuration lies in close proximity to the nanoribbon (~ 2.4 Å), and because the strong dipole potential decays slowly in space ($V \sim d/r^2$), the degeneracy of the up- and down-spin states on the separate ZNGR edges is now completely broken (cf. bottom of Figure 3), and a significant re-ordering of electronic bands occurs, as shown in Figure 5. This effect is most pronounced near the Fermi level where the band gap of the spin-down state is reduced to 0.33 eV while the band gap of the spin-up state is nearly doubled to a value of 0.56 eV. We should mention at this point that we were able to reproduce these same effects using other different DFT functionals and approaches (described further in the Theoretical Methods section), indicating that the spin-switching behavior in nanoribbon-spiropyran is rather robust and insensitive to the specific functional or pseudopotential used. It is also important to point out that our calculations were strongly motivated from available spiropyran materials currently accessible from our own previous experimental efforts,[24] and that further modifications of the spiropyran unit could lead to even larger spin-polarized effects. In particular, even with our relatively small ZGNR (which has a large band gap of 0.45 eV), we were able to achieve a nearly 50% change in band gap between the up- and down-spin states using our unmodified, "as given" spiropyran-embedded polymer. Further chemical functionalization with other electron donor and acceptor groups[5,30-32] to magnify the dipole of the open spiropyran in conjunction with wider ZGNRs (which have much smaller intrinsic band gaps) would even further reduce this band gap. Nevertheless, our calculations clearly demonstrate that the net effect of a switchable polar spiropyran material is to alter the spin states in ZGNR as though it were under an external electric field. However, in contrast to an isolated ZGNR, the effects shown here are *intrinsic* to our nanoribbon-spiropyran hybrid structure, eliminating the need for external electric fields to induce spin-switching.

Finally, in these last sections we consider the reversible opto-mechanical transformations of the nanoribbon-spiropyran structure as driven by both mechanical deformation and by optoelectronic excitation. As mentioned in the introduction, we have previously carried out dynamic-mechanical-analysis experiments on a spiropyran-embedded polymer to probe the mechanochromic properties of these materials.[24] We also note that we have previously carried out opto-mechanical calculations on this system (we did not, however, perform calculations on the dipole moment, which we discuss later), so the mechanical deformation and optoelectronic calculations described in this section are briefly reviewed here again to provide the necessary background for understanding the dipole-moment transitions in this system. In order to quantitatively predict the specific opto-mechanical conditions that induce spin-symmetry breaking in our nanoribbon-spiropyran hybrid system, we carried out a series of force-constrained DFT and time-dependent DFT calculations. First, to simulate an applied stress, the initial equilibrium geometry for a closed spiropyran unit tethered with pendant poly($\varepsilon$-caprolactone) polymer chains was optimized with DFT. Starting with this equilibrium geometry, an applied external stress was obtained by gradually increasing the distance between the two terminal methyl groups (denoted by arrows in Figure 6a) in small increments of 0.1 Å up to a final molecular elongation of 60%. Throughout this entire procedure, all of the other unconstrained internal coordinates were fully optimized in order to minimize the total strain energy. The resulting energy curve is depicted in Figure 6b, which shows a monotonic increase in energy as the polymer is stretched until an abrupt transition occurs at 39% elongation. At this transition point, the C-O spiro bond suddenly ruptures, and the energy sharply decreases from 2.7 to 1.6 eV as the spiropyran has now relaxed into the open charge-separated zwitterionic merocyanine form. It is remarkable to point out that our calculations naturally predict the C-O spiro bond to be the weakest in the entire spiropyran polymer, even though we have not constrained this particular bond-energy at all in our first-principles calculations. After the weak C-O spiro bond has broken, the energy for elongations greater than 39% are associated with stretching the strong molecular bonds along the open zwitterionic backbone.

Next, in parallel with the force-constrained DFT optimizations, we also carried out excited-state calculations to understand the photochromic properties of the nanoribbon-spiropyran polymer. As described in our previous work, *reversible* photoswitching of a spiropyran-embedded polymer is experimentally realizable since light of *different* wavelengths can initiate both the forward and reverse isomerizations in these materials. In particular, real-time monitoring of a spiropyran-polymer film gives a dramatic demonstration of a reversible cyclic response to light irradiation.[24] In order to evaluate the photoresponse in this system, we carried out several time-dependent DFT calculations at each of the force-constrained polymer geometries. In Figure 6c, we plot the excitation wavelength having the strongest absorption maximum along the closed-spiropyran → open-merocyanine reaction path (further technical details are given in the Theoretical Methods section). For molecular elongations less than 39%, the closed spiropyran configuration is the most stable, and Figure 6c shows that it only absorbs in the UV with a maximum absorbance at 250 nm. However, upon cleavage of the weak C-O spiro bond, the open merocyanine structure shows a strong $S_0 \rightarrow S_1$ absorption maximum (450 nm) that is significantly red-shifted due to the larger conjugated $\pi$-electron system, confirming previous experimental observations of a strong photochromic shift in the polymer. To further characterize this dramatic change in electronic character, we also performed new calculations of the electric dipole moment in Figure 6d along the spiropyran → merocyanine reaction path. In the closed-spiropyran region, the calculated dipole moment has a value of 7.5 D that begins to change discontinuously in the same energy region where the C-O spiro bond is broken and where the polymer begins to strongly absorb at 450 nm. Beyond this transition point, the dipole moment almost doubles to a value of 11.5 D due to increasing polarization of the C-O bond as the polymer is stretched. As discussed previously, it is at this transition point that the electric dipole is now strong enough to create a potential energy difference for completely breaking the spin symmetry of the nanoribbon-spiropyran hybrid system. Our first-principles calculations give detailed mechanistic insight into this evolution of the dipole moment, and they further predict the resulting spin-symmetry breaking in the nanoribbon-spiropyran system to occur at 39% elongation, or at the 250 nm/450 nm crossover point.

**Conclusions**

In this study, we have shown that a spiropyran-embedded polymer noncovalently deposited across a nanoribbon can collectively function as a dual opto-mechanical switch for modulating its own spin-polarization. Using both first-principles DFT and time-dependent DFT, we demonstrate that this unique switching of spin states is triggered by the *reversible* isomerization of a closed-configuration spiropyran to a charge-separated merocyanine form, which alters the electrostatic environment of the nanoribbon. Specifically, we show that upon mechanical stress (39% elongation) or photoabsorption (between 250 nm and 450 nm), a large dipole field is created, generating a potential energy difference across the nanoribbon edges that completely breaks their spin symmetry. Using an unmodified, experimentally-accessible spiropyran material, we were able to achieve a nearly 50% change in band gap between the up- and down-spin states throughout the nanoribbon substrate. More importantly, the mechanophoric effects described here are *intrinsic* to our hybrid system, allowing us to reversibly modulate the spintronic properties of zigzag nanoribbons via noninvasive optical and mechanical stimuli without the need for large external electric fields.

Looking forward, it would be extremely interesting to postulate how one can further optimize this hybrid system for enhanced spintronic transport or how to incorporate these systems in future nanoelectronic applications. As mentioned previously, while our calculations were motivated from our own experimentally-available spiropyran materials, it is very likely that further modifications of the spiropyran unit would lead to even larger spin-polarized effects. In particular, further chemical functionalization via other strong electron donor or acceptor groups of the spiropyran in conjunction with wider nanoribbons would even further reduce the band gap towards a half-metallic state. We are currently exploring these other options in parallel with experimental efforts to understand their effect on spin-polarized current within a non-equilbrium Green's function formalism. Other potential extensions of our work include the possibility of using the spiropyran-embedded polymer (or even the isolated spiropyran unit itself for molecular electronics applications) within mechanically controllable break junction (MCBJ) experiments.[33,34] For instance, it is conceivable that upon stretching a spiropyran unit suspended in a MCBJ apparatus, one can monitor both the nanomechanical and photoinduced electronic transport in a nanoribbon substrate in a highly-controlled environment. Lastly, the use of spin-polarized nanostructures that can be modulated by external stimuli naturally lends itself to applications in quantum computing and information. While many molecular logic gates reported in the literature are not resettable, the reversible spintronic properties inherent to the nanoribbon-spiropyran material allow the possibility of using these structures as write-read-erasable units for quantum information. Specifically, the nanoribbon-spiropyran hybrid system can function as a resettable and multi-readout molecular-logic sensor by using opto-mechanical stimuli as inputs and the spin-polarized current induced in the nanoribbon substrate as the measured output. Using external stimuli to coherently control electronic spins and modulate their coupling with the environment is of great importance in quantum computing and error-propagation. As a result, we anticipate that the spintronic effects presented here are applicable to opto-mechanical switches in other nanostructures, and may provide further insight into substrate/environment effects in quantum-coupled systems of increasing complexity.

**Theoretical Methods:**

Our results for the nanoribbon-spiropyran hybrid material are based on first-principles density functional theory (DFT) calculations as implemented in the Quantum ESPRESSO suite of computer codes.[35] Spin-polarized, generalized gradient approximation (GGA) calculations with the Perdew-Burke-Ernzerhof (PBE) functional[36] were performed with ultrasoft pseudopotentials under periodic boundary conditions. In order to include the van der Waals interaction[26] between the spiropyran and the nanoribbon, we also added dispersion effects in all of our calculations using the DFT-D approach by Grimme.[25,26] Within the DFT-D approach, an atomic pairwise dispersion correction is added to the Kohn-Sham part of the total energy ($E_{\text{KS-DFT}}$) as

$$E_{\text{DFT-D}} = E_{\text{KS-DFT}} + E_{\text{disp}},$$

where $E_{\text{disp}}$ is given by

$$E_{\text{disp}} = -s_6 \sum_{i=1}^{N_{\text{at}}-1} \sum_{j=i+1}^{N_{\text{at}}} \sum_{\mathbf{g}} f_{\text{damp}}\left(R_{ij,\mathbf{g}}\right) \frac{C_6^{ij}}{R_{ij,\mathbf{g}}^6}.$$

Here, the summation is over all atom pairs $i$ and $j$, and over all $\mathbf{g}$ lattice vectors with the exclusion of the $i = j$ contribution when $\mathbf{g} = 0$ (this restriction prevents atomic self-interaction in the reference cell). The parameter $C_6^{ij}$ is the dispersion coefficient for atom pairs $i$ and $j$, calculated as the geometric mean of the atomic dispersion coefficients:

$$C_6^{ij} = \sqrt{C_6^i C_6^j}.$$

The $s_6$ parameter is a global scaling factor that is specific to the adopted DFT method ($s_6 = 0.75$ for PBE), and $R_{ij,\mathbf{g}}$ is the interatomic distance between atom $i$ in the reference cell and $j$ in the neighboring cell at distance $|\mathbf{g}|$. A cutoff distance of 50.0 Bohrs was used to truncate the lattice summation, which corresponds to an estimated error of less than 0.02 kJ/mol on cohesive energies, as determined by previous studies.[26,37] In order to avoid near-singularities for small interatomic distances, the damping function has the form

$$f_{\text{damp}}\left(R_{ij,\mathbf{g}}\right) = \frac{1}{1+\exp\left[-d\left(R_{ij,\mathbf{g}}/R_{\text{vdW}}-1\right)\right]},$$

where $R_{\text{vdW}}$ is the sum of atomic van der Waals radii $\left(R_{\text{vdW}} = R_{\text{vdW}}^{i} + R_{\text{vdW}}^{j}\right)$, and $d$ controls the steepness of the damping function. To prevent spurious dipole interactions between adjacent spiropyran chromophores in repeated supercells, the vacuum along the axes perpendicular to the nanoribbon ($z$-axis in Figure 2) was set to 100 Bohrs. Unconstrained geometry optimizations of both the ions and the supercell were carried out with a kinetic-energy cutoff of 25 Ry (~ 340 eV), a charge density cutoff of 200 Ry (~ 2721 eV), and a 10 × 1 × 1 Monkhorst-Pack grid uniformly distributed along the periodic axis of the nanoribbon. From the converged DFT-D optimized geometries, we then computed the spin-polarized band structure using a dense 200 × 1 × 1 Monkhorst-Pack grid along the 1D Brillouin zone of the nanoribbon. To further verify our DFT-D energies and spin-polarized band structures, we also performed benchmark comparisons using projected augmented wave (PAW) pseudopotentials with the unmodified PBE functional (without dispersion) as implemented in the Vienna *Ab initio* Simulation Package (VASP).[38] Based on these benchmark calculations, we found there was no significant difference between the spin-polarized band structures based on PBE-D and those based on the unmodified PBE functional, indicating that the spin-switching behavior in nanoribbon-spiropyran is rather robust and insensitive to the specific functional or pseudopotential used. It should be mentioned, however, that both the PBE-D and PBE band structures are still expected to underestimate bandgaps for the nanoribbon structure (due to self-interaction errors inherent to semi-local GGA functionals that over-delocalize electrons in extended systems[39-41]); therefore, our calculated values should be considered as lower bounds to the experimental band gap. We are currently exploring the use of more computationally-intensive range-separated hybrid functionals[42-48] in these fairly large nanoribbon-spiropyran systems (over 100 spin-polarized atoms in a sizeable vacuum space), which we save for future work.

For the time-dependent DFT (TD-DFT) calculations, we used a locally-modified version of the Gaussian 09 code to compute electronic excitation energies as a function of mechanical stress. The initial equilibrium geometry for the closed spiropyran polymer was optimized using the recent LC-BLYP functional (range-separation parameter set to $\mu = 0.31$)[46,47] with the 6-31G(d,p) basis set. From this initial equilibrium geometry, an applied stress was calculated by gradually increasing the distance between the two terminal methyl carbon atoms (cf. Figure 6a) while optimizing all of the other internal coordinates to minimize the total energy of the strained system. At each of these relaxed geometries, both single-point energies/dipoles and TD-DFT excitation energies were carried out with a diffuse 6-31+G(d,p) basis set. In each of the TD-DFT calculations, the lowest 10 singlet electronic excitations were computed, and the oscillator strengths were obtained from the converged eigenvectors of the TD-DFT linear-response equations. Electric dipole moments for the ground state were obtained by evaluating expectation values of the dipole operator from the orbital-relaxed LC-BLYP density matrix.

*Acknowledgment.* This research was supported in part by the National Science Foundation through TeraGrid resources (Grant No. TG-CHE1000066N) provided by the National Center for Supercomputing Applications. Funding for this effort was provided by the Laboratory Directed Research and Development (LDRD) program at Sandia National Laboratories, a multiprogram laboratory operated by Sandia Corporation, a Lockheed Martin Company, for the United States Department of Energy under contract DE-AC04-94AL85000.

**Figure captions**

**Figure 1.** Molecular structures of a closed-configuration spiropyran and its charge-separated open merocyanine form. The open structure can be formed by photoexcitation with UV irradiation or through an externally-applied mechanical stress. The isomerization process is reversible upon irradiation with visible light, which restores the closed spiropyran form.

**Figure 2.** Equilibrium configuration of an open merocyanine unit noncovalently deposited on a free-standing ZGNR. The merocyanine lies relatively flat on the ZGNR with a separation distance of 2.4 Å (measured from the lowest atom of merocyanine to the ZGNR surface). The ZGNR repeats periodically along the $x$-axis with a large vacuum space of 100 Bohrs along the $y$- and $z$-axes.

**Figure 3.** Top-down views of the open and closed spiroypyran forms deposited on free-standing ZGNRs. The width of the unit cell is indicated by the dashed box on the left. Upon UV irradiation or applied stress, the closed-configuration isomerizes to an open form that has a large dipole moment (11.5 Debye). The resulting dipole perturbs the electrostatic environment of the nanoribbon, leaving localized edge states of only one spin orientation, as schematically depicted at the bottom of each panel.

**Figure 4.** Isosurface of the charge-density difference between the up- and down-spin states for the open merocyanine unit deposited on a ZGNR. The periodic repeat units have been omitted for clarity.

**Figure 5.** Electronic band structures for an isolated ZGNR, a closed spiropyran form on ZGNR, and an open merocyanine-ZGNR configuration. The unmodified ZGNR has up- and down-spin states that are energetically degenerate and, therefore, overlap. For the closed spiropyran form on ZGNR, the symmetry of the up- and down-spin states has broken, and a slight widening of the spin-up band gap (relative to the spin-down band gap) can be seen near the Fermi level (which lies at ~ -3.7 eV in each plot). In the last panel, the open merocyanine form has a sizeable dipole moment that creates a large potential energy difference across the ZGNR width, leaving electronic states with only one spin orientation (down-spin) near the Fermi level.

**Figure 6.** (a) Molecular structure of a closed spiropyran unit tethered with pendant poly(ε-caprolactone) polymer chains. An external mechanical stress was applied across this polymer unit by gradually increasing the distance between the methyl units (denoted by arrows) while optimizing all other internal coordinates. (b) Force-constrained potential energy curve for mechanochemical switching of closed spiropyran to the open merocyanine form. A sharp transition between the spiropyran and merocyanine forms occurs at 39% elongation length. (c) Absorption wavelength for photochromic switching between closed spiropyran and open merocyanine configurations. The closed spiropyran form absorbs in the UV (< 39 % elongation), while the open merocyanine form strongly absorbs in the visible (> 39% elongation). (d) Electric dipole moment as a function of molecular elongation. When the closed spiropyran polymer is stretched, the dipole moment nearly doubles as the isomerization proceeds towards the zwitterionic, open-merocyanine configuration.

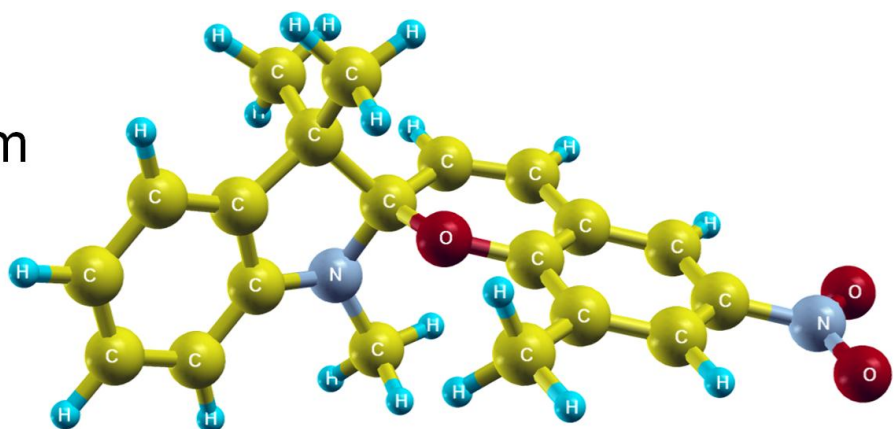
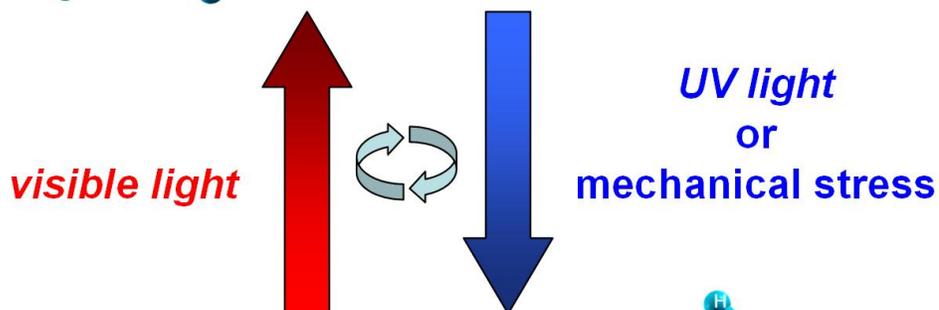
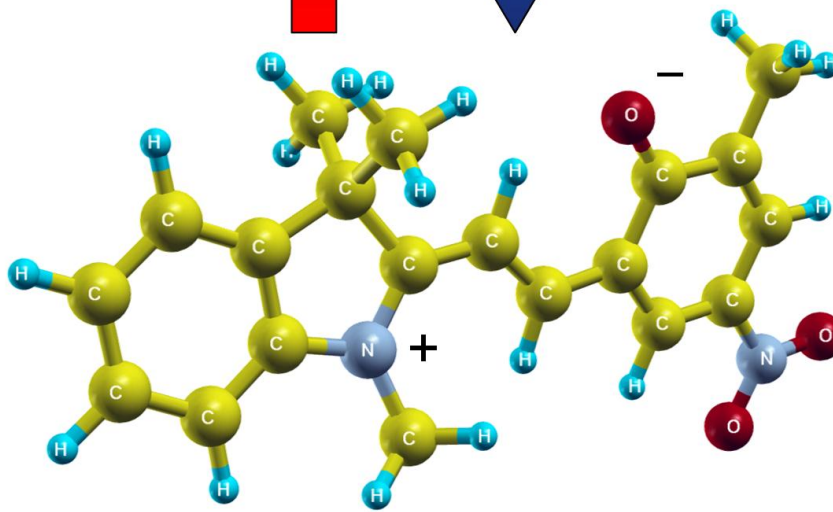

**Figure 1**

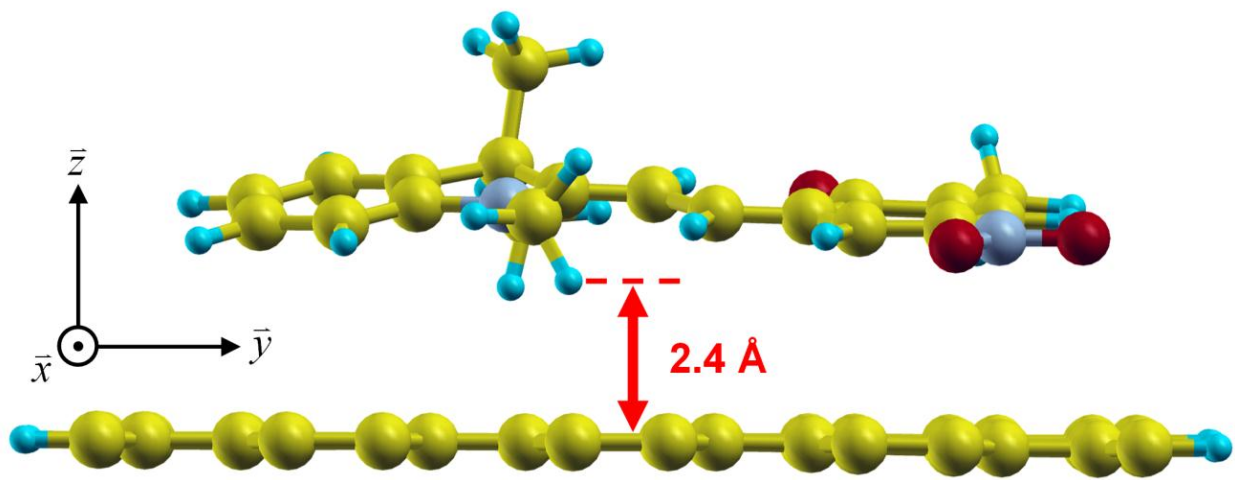

**Figure 2**

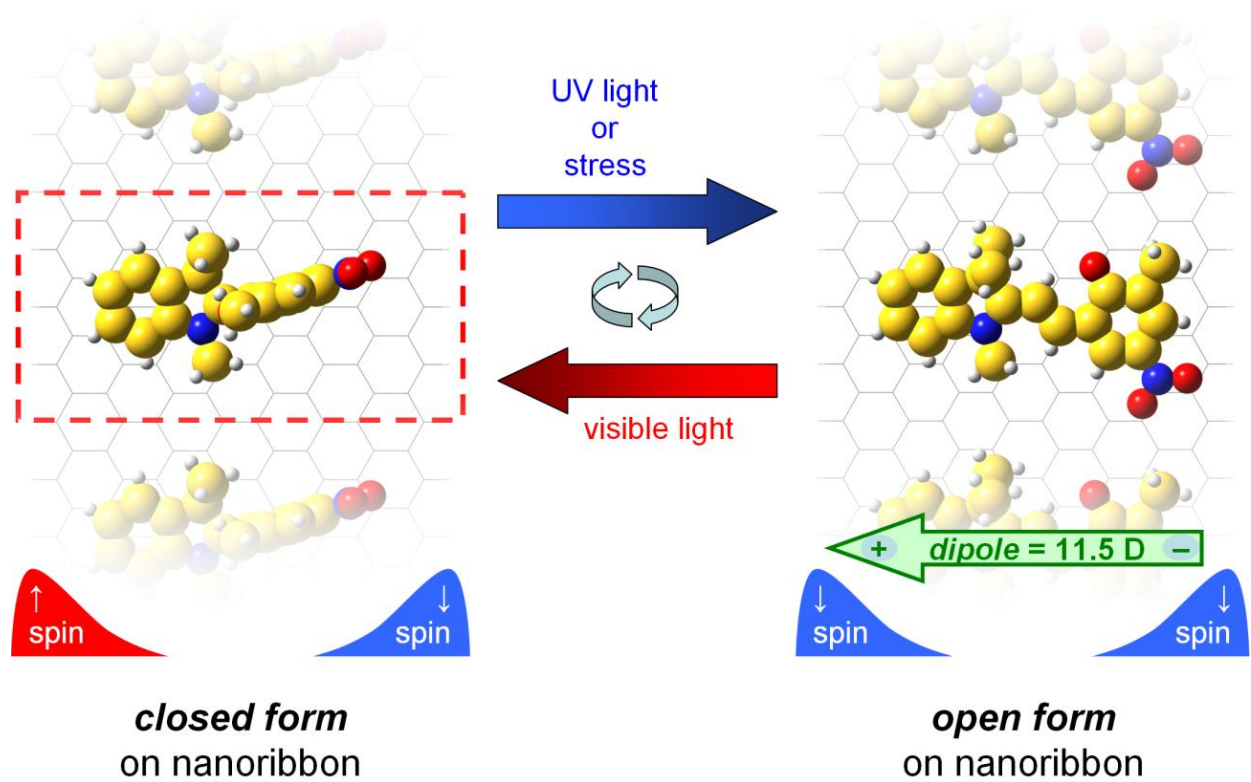

**Figure 3**

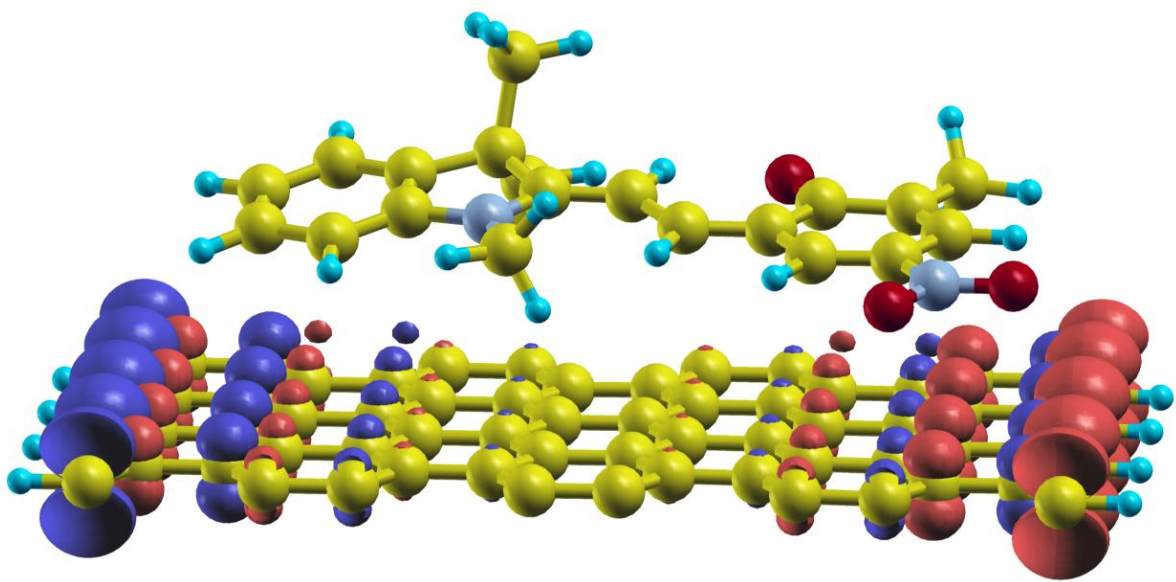

**Figure 4**

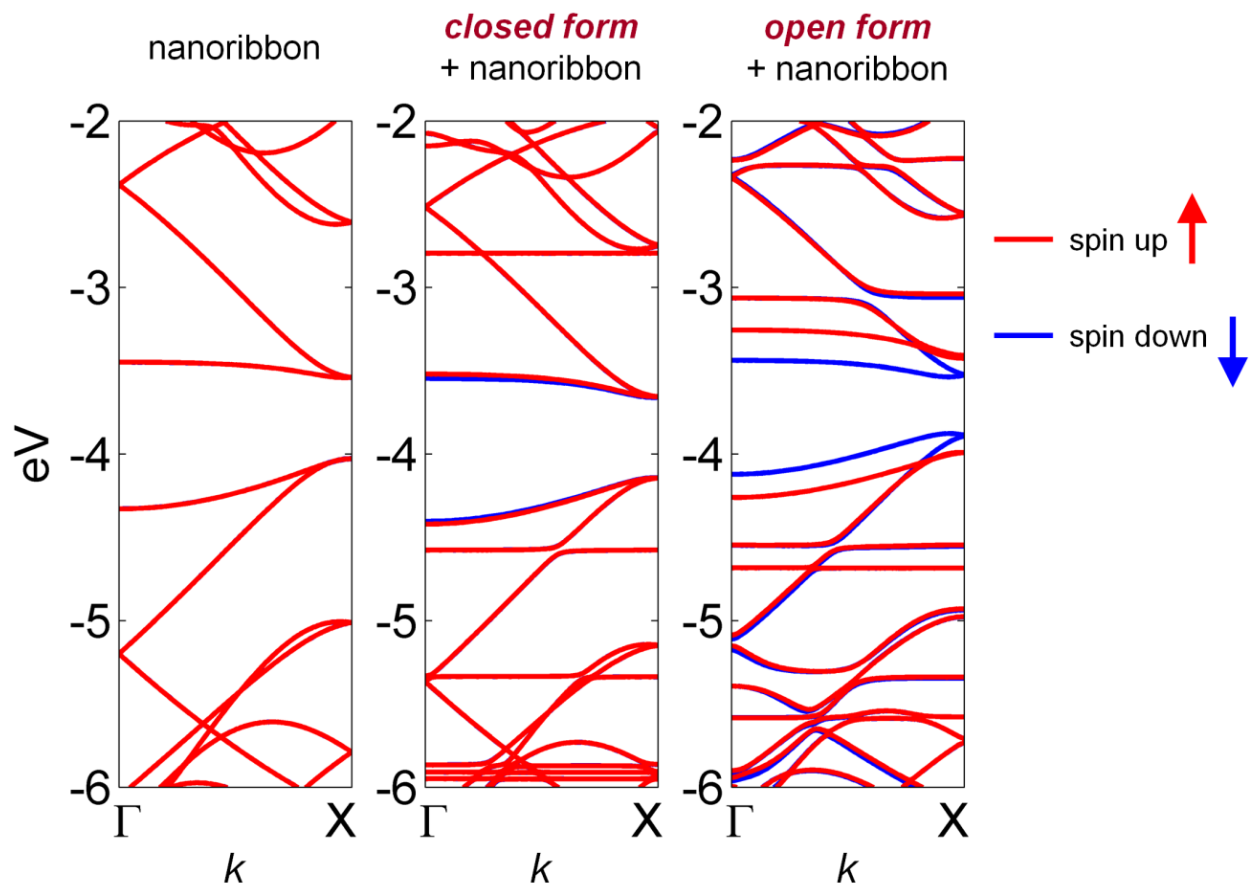

**Figure 5**

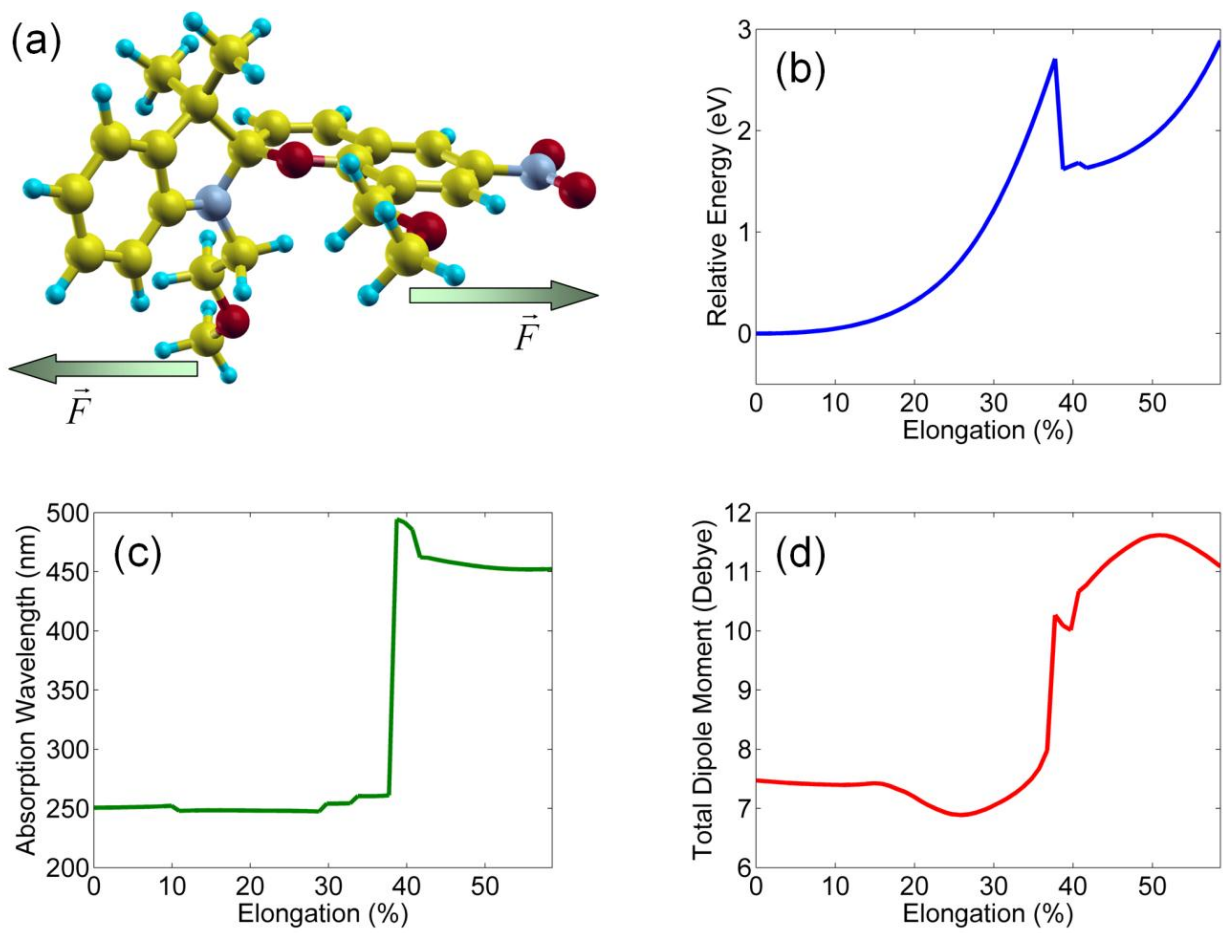

**Figure 6**